\documentclass[fleqn,usenatbib]{mnras}

\usepackage{newtxtext,newtxmath,multirow}

\usepackage{pdflscape}
\usepackage{booktabs}     
\usepackage{multirow}     
\usepackage{makecell}     
\usepackage{tabularx}     
\usepackage{siunitx}      
\usepackage{rotating}     
\usepackage{graphicx}     
\usepackage{caption}      
\usepackage{array}        
\usepackage{lineno}
\usepackage{ulem}
\usepackage{mathtools}

\usepackage[T1]{fontenc}

\DeclareRobustCommand{\VAN}[3]{#2}
\let\VANthebibliography\thebibliography
\def\thebibliography{\DeclareRobustCommand{\VAN}[3]{##3}\VANthebibliography}

\usepackage{graphicx}	
\usepackage{amsmath}	
\usepackage{lscape}

\title[DTW qualification of transitional pulsar candidates]{Assessing transitional pulsar candidates through a quantitative estimation of their light-curve similarity}

\author[D. F. Torres et al.]{
D. F. Torres$^{1,2,3}$\thanks{dtorres@ice.csic.es},
C. R. Garc\'ia$^{1,2,4}$\thanks{crodriguez@ice.csic.es},
F. Coti Zelati$^{1,2}$, 
D. de Martino$^{5}$, 
G. Illiano$^{1,2,6}$, 
A. Papitto$^{7}$, 
\newauthor
A. Patruno$^{1,2}$, \&
M. C. Baglio$^{6}$
\\
$^{1}$ Institute of Space Sciences (ICE, CSIC), Campus UAB, Carrer 
de Can Magrans s/n, Barcelona 08193, Spain \\
$^{2}$ Institut d'Estudis Espacials de Catalunya (IEEC), Barcelona 08860, Spain \\ 
$^{3}$ Instituci\'o Catalana de Recerca i Estudis Avan\c{c}ats (ICREA), Barcelona, Spain \\
$^{4}$ Institute for Research in Technology (IIT), Universidad Pontificia Comillas, Madrid, Spain \\
$^{5}$ INAF-Osservatorio Astronomico di Capodimonte, Salita Moiariello 16, I-80131, Naples, Italy \\
$^{6}$ INAF-Osservatorio Astronomico di Brera, Via Bianchi 46, I-23807 Merate (LC), Italy \\
$^{7}$ INAF-Osservatorio Astronomico di Roma, via di Frascati 33, I-00078, Monteporzio Catone, Roma, Italy
       }

\date{Accepted XXX. Received YYY; in original form ZZZ}

\pubyear{\the\year{}}

\begin{document}
\label{firstpage}
\pagerange{\pageref{firstpage}--\pageref{lastpage}}
\maketitle

\begin{abstract}
Transitional millisecond pulsars are often identified through X-ray light-curve morphology, especially by visual comparison with the prototype PSR J1023+0038. Such classifications remain qualitative and lack a calibrated notion of how much variability is already present within the prototype itself. We introduce a Dynamic Time Warping (DTW) framework that compares light curves through moving-window similarity distributions and calibrates candidate-prototype distances against the internal heterogeneity of PSR J1023+0038. In this formulation, a source is not simply asked to resemble one prototype observation, but to remain compatible with the range of morphologies exhibited by the prototype across observations. We apply the method to confirmed and candidate tMSPs and to a control sample of non-tMSP systems. The resulting similarity measure provides a reproducible ranking for tMSP-like behaviour and is therefore useful for candidate triage and archival searches. At the same time, the control sample shows that short-timescale accretion variability can partly mimic tMSP-like morphology, demonstrating that light-curve shape alone is a screening diagnostic rather than a definitive class identifier.
\end{abstract}


\section{Introduction} 

Transitional millisecond pulsars (tMSPs) are compact neutron-star binaries that alternate between an accretion-powered state and a rotation-powered radio-pulsar state, thereby providing a direct observational link between low-mass X-ray binaries and recycled pulsars \citep[see, e.g.,][]{Archibald2009, Papitto2013, Papitto_2022}. 
In the sub-luminous accreting regime, their X-ray emission exhibits a distinctive combination of (i) pronounced stochastic variability on seconds-to-minutes timescales and (ii) recurrent switching between quasi-stable luminosity modes, commonly labelled high, low, and flaring modes \citep[see, e.g.,][]{Linares_2014,Bogdanov2015}. 
Prototypical systems such as PSR~J1023+0038 have established this mode-switching phenomenology as the practical signature for identifying possible members of the class.

Candidate selection, however, still rests on qualitative pattern recognition: light curves are commonly deemed ``tMSP-like'' by visual comparison with prototypical sources. 
This approach is effective for choosing targets for additional observations, but it is intrinsically subjective and does not provide a quantitative measure of resemblance, nor an explicit calibration against the internal heterogeneity of the prototype itself. 
In addition, differences in observation length, cadence, and stochastic variability can bias purely qualitative judgments. 
The issue is therefore not only one of formalizing visual resemblance, but of enabling reproducible candidate ranking across heterogeneous observations, so that morphology can be used as an operational first-pass criterion in source selection and archive searches.

Here we address these limitations by introducing a quantitative, morphology-based comparison of X-ray light curves using Dynamic Time Warping \citep[DTW, see, e.g.,][]{Berndt_1994}.
The DTW approach is broadly utilized across various fields, from speech recognition to analyze market trends.
DTW has also been recently used in astrophysics.
Examples include the similarity between light curves of gamma-ray bursts \citep{Zhang_2016}, the reconstruction of gravitational wave signals from core-collapse supernovae \citep{Suvorova2019}, and the assessment of solar wind time series \citep{Samara2022}. \citet{Vohl2024} have employed an unsupervised method involving DTW to categorize radio pulsars by profile shape similarity using graph theory.
In \citet{DTW_I}, we use DTW to analyze the population of gamma-ray-emitting pulsars detected by Fermi-LAT 3PC,
quantitatively showing that pulsars with similar light curves can have different physical and spectral properties.

\section{Source selection and their X-ray light curves}
\label{sec: preprocessing_light_curves}

The light curves analyzed here include confirmed and candidate tMSPs: PSR J1023+0038 (also referred to as the prototypical source or J1023), XSS~J12270-4859, IGR~J18245-2452, and the candidates 1RXS~J154439.4-112820, CXOU~J110926.4-650224.
To assess the methodology's discriminative power, we also include a control sample of systems (formed by accreting millisecond pulsars (AMSPs) and cataclysmic variables (CVs)) consisting of: PBC~J0801.2-4625, IGR~J12123-5802, SWIFT~J0746-1611, IGR~J17511-3057, and SAX~J1808.4-3658.
Each system is briefly introduced in Appendix \ref{appendix: src_sample} together with details regarding the observations used to characterize them.

The preparation of the X-ray light curves for the application of the methodology is described in Appendix \ref{appendix: preprocessing_light_curves}.
The two more relevant steps are the identification of temporal discontinuities and the removal of flares.
We extract light curve segments with a minimum temporal length of 1h (3600 s, or 120 bins of 30 s, in which all light curves are shown). 
The size of the temporal bin and the minimum duration to be considered a light curve segment can be chosen according to data.
The aim of these choices is to avoid comparisons with too short light curves which may be inappropriate to define the systems' mode; and to define a duration of a bin that can be equally used in all systems to be compared.

Since the aim of the present method is to compare the recurrent mode-switching morphology of the sub-luminous disc state, we exclude flaring intervals before applying DTW. 
Otherwise their sporadic occurrence, broad amplitude range, and source-to-source variability would dominate any distance.
When existing, flares of confirmed and candidate tMSPs are identified and removed using the count-rate distribution.
A threshold is set at three standard deviations above the peak of the highest-count distribution.
This threshold is estimated assuming an approximate Gaussian distribution of the high mode. 
Bins with counts exceeding this threshold are classified as flaring modes and are excluded.
After flare removal, we require that the remaining light curve segments meet the minimum duration requirement of 3600~s.

Each resulting light curve is normalized using the min-max scaler (see Appendix \ref{appendix: preprocessing_light_curves}, Eq. \ref{eq: min-max_scaler}).
This ensures that each light curve is evaluated mainly based on its morphology, regardless of flux variations, and that its values are adjusted between 0 and 1.
Note that to prevent distorting the light curve shapes, the min-max scaler is applied collectively to all those segments originating from the same observation.
Table~\ref{tab: sample_summary} shows the number of observational campaigns available for each system, along with the resulting number of light curve segments obtained.
Appendix \ref{appendix: preprocessing_light_curves} shows all the light curves used in this study.

\section{Quantifying light curve similarity}

\subsection{Basics of dynamic time warping}
\label{sec: basics_dtw}

DTW finds the optimal alignment between two-time series by allowing flexible comparisons while preserving their structural integrity.
Consider two time series $X_{\mathrm{\mathrm{LC}}}$ and $Y_{\mathrm{\mathrm{LC}}}$ of length $n$ and $m$:
\begin{equation}  
    \begin{aligned}
        X_{\mathrm{\mathrm{LC}}} &= [x_1, x_2, \dots, x_i, \dots, x_n] \\
        Y_{\mathrm{\mathrm{LC}}} &= [y_1, y_2, \dots, y_j, \dots, y_m]~.
    \end{aligned}
    \label{eq: time_series}
\end{equation}
The indices $(i,j)$ in Eq.~(\ref{eq: time_series}) run along the elements of each time series, respectively ($i=1, \dots, n$, $j=1, \dots, m$).
To be considered a warping path, i.e., a set of connected pairs of points $(x_i, y_j)$ covering all the elements of both series in Eq.~(\ref{eq: time_series}), we shall require the warping path conditions \citep[see, e.g., ][]{Berndt_1994}: the usual boundary ($i_{1}=1$, $j_{1}=1$, and $i_{k}=n$, $j_{k}=m$), monotonicity ($i_{k-1} \leq i_{k} $ and $j_{k-1} \leq j_{k} $), and continuity ($i_{k}-i_{k-1} \leq 1 $ and $j_{k}-j_{k-1} \leq 1 $) conditions.
Taking these conditions into account, the DTW is computed\footnote{The computation follows the implementation in the \texttt{dtaidistance} Python package \citep{dtaidistance}}. as \citep[see, e.g.,][]{ShokoohiYekta2014GeneralizingDT}:
\begin{eqnarray}
\text{DTW}(X_{\mathrm{\mathrm{LC}}}, Y_{\mathrm{\mathrm{LC}}}) = 
\min_{W(X_{\mathrm{\mathrm{LC}}}, Y_{\mathrm{\mathrm{LC}}})\in \mathcal{A}}
\Big(\sum_{(i,j)\in W}d(x_{i},y_{j})^{2}\Big)^{\frac{1}{2}}
,\label{eq: DTW_optimized}
\end{eqnarray}
where $W$ represents an explored warping path that pertains to the set $\mathcal{A}$ of all paths fulfilling the  warping path conditions. 
The DTW value from Eq.~(\ref{eq: DTW_optimized}) is the cost of the optimal warping path ($W_{\mathrm{o}}$), that is, it is the minimum value obtained after adding the distance $d$ between all pairs $(i,j)$ of $W_{\mathrm{o}}$.
Besides the warping path conditions, we can also impose a global constraint.
Two typical global constraints are the Sakoe-Chiba band \citep{Sakoe1978} and the Itakura parallelogram \citep{Itakura_75}; see, e.g., \cite{SakoeVsItakura} for a discussion.
A comparison of these constraints, together with other alternatives and the moving-window process adopted here, is discussed in Appendix~\ref{appendix: alternative_dtw_constraints}.

Here we adopt a Sakoe-Chiba band with a warping window maximum size ($w_w$).
The parameter $w_w$ limits pathological warpings in which few elements of $X_{\mathrm{\mathrm{LC}}}$ are aligned with many elements of $Y_{\mathrm{\mathrm{LC}}}$ to minimize the cost.
Larger values of $w_w$ allow more flexible and potentially degenerate alignments, which can artificially increase similarity, while smaller values enforce stricter correspondences.
See additional discussion in Appendix~\ref{appendix: hyp}.

\subsection{Moving window sets}
\label{sec: methodology_dtw_tmsp}

We apply a moving window to divide each light curve into smaller, partially overlapping segments.
We call these segments, windows ($w$). Appendix \ref{appendix: mov-win} gives an example.
We express the size of each light curve ($N$) in bins (length $n$ or $m$ in Eq.~(\ref{eq: time_series})); therefore, the implementation depends on the window size ($w_s$) and the window step size ($s_s$), both also measured in bins. 
See Appendix \ref{appendix: hyp} for further discussion.
The total number of windows ($N_w$) for a given light curve is calculated as\footnote{The $\lfloor \cdot \rfloor$ operation, called floor division, guarantees that the result is a whole number, in this case, by rounding down to the nearest one. }:
\begin{eqnarray}
N_{w}=\left \lfloor \frac{N-w_{s}}{s_{s}} \right \rfloor +1.
\label{eq: number_windows}
\end{eqnarray}
After applying Eq.~(\ref{eq: number_windows}), any ignored portion of the light curve has less than $w_s$ bins.
Therefore, the times series provided in Eq.~(\ref{eq: time_series}), can be represented using the moving windows as 
\begin{equation}  
    \begin{aligned}
        X_{\mathrm{\mathrm{LC}}} &= \{w_1, w_2, \dots, w_i, \dots, w_{N_{w}(n)}\} \\
        Y_{\mathrm{\mathrm{LC}}} &= \{w_1, w_2, \dots, w_j, \dots, w_{N_{w}(m)}\}~.
    \end{aligned}
    \label{eq: time_series_windows}
\end{equation}

\subsection{Similarity comparison methodology based on DTW}
\label{sec: similarity_DTW}

We apply DTW using Eq.~(\ref{eq: DTW_optimized}) for each pair of windows $w$ in Eq.~(\ref{eq: time_series_windows}). 
This generates a set of values $d_{ij}=\text{DTW}(w_i , w_j)$.
Examples of such matches are given in Appendix \ref{appendix: win-matching}.
For each $w_i \in X_{\mathrm{\mathrm{LC}}}$, we only keep the minimum DTW value calculated (i.e., that corresponding to the most similar window) among all comparisons with windows $w_j \in Y_{\mathrm{\mathrm{LC}}}$, referred to as $d_{i}^{X_{\mathrm{\mathrm{LC}}} \rightarrow Y_{\mathrm{\mathrm{LC}}}} =\min_{1\leq j \leq N_{w}(n)} d_{ij}$.
Similarly, for each $w_j \in Y_{\mathrm{\mathrm{LC}}}$, we only keep the minimum DTW value calculated among all comparisons with $w_i \in X_{\mathrm{\mathrm{LC}}}$, referred to as $d_{j}^{Y_{\mathrm{\mathrm{LC}}} \rightarrow X_{\mathrm{\mathrm{LC}}}} =\min_{1\leq i \leq N_{w}(m)} d_{ij}$.
We then define the pooled set of these minimal distances as\footnote{The $\biguplus$ operator denotes the multiset union, which preserves the multiplicity of all elements rather than removing duplicate values.}:
\begin{equation}  
    \begin{aligned}
        & \mathcal{D}_{XY} = \{  d_{i}^{X_{\mathrm{\mathrm{LC}}} \rightarrow Y_{\mathrm{\mathrm{LC}}}} \}^{N_{w}(n)}_{i=1} \biguplus  \{d_{j}^{Y_{\mathrm{\mathrm{LC}}} \rightarrow X_{\mathrm{LC}}} \}^{N_{w}(m)}_{j=1}~.
    \end{aligned}
    \label{eq: DTW_similarity_methodology}
\end{equation}
We shall consider a measure of similarity based on the inverse of Eq.~(\ref{eq: DTW_optimized}), $s=\text{DTW}^{-1}$.
Thus, more similar comparisons will produce higher values of $s$. 
Next we treat the values in $\mathcal{D}$ using Eq.~(\ref{eq: DTW_similarity_methodology}) as an empirical cumulative distribution function (ECDF).
The ECDF of the set of minimum DTW between two light curves will be denoted as $\widehat{F}_{X Y}(s)$.

\subsection{Similarity based on ECDFs of the minimum DTW values}
\label{sec: similarity_ECDFs}

The similarity of a system $Z$ to the prototype system (J1023) is to be characterised by an ECDF, which we shall refer to as $\widehat{F}_{Z}(s)$.
This ECDF is constructed from the pooled set of DTW-derived values obtained by comparing the light curves of $Z$ with each available prototype light curve of J1023, following Eq.~(\ref{eq: DTW_similarity_methodology}).

Let $D_Z = \{ d_1, \dots, d_K \}$ denote the pooled sample of DTW distances used to construct $\widehat{F}_{Z}(s)$. 
Explicitly, if $\mathcal{D}_{Z_j X_i}$ denotes the collection of window-window minimal DTW distances obtained from the comparison between the $j$-th light curve of $Z$ and the $i$-th prototype light curve of J1023, then $D_Z = \biguplus_{j,i} \mathcal{D}_{Z_j X_i}$, according to Eq.~(\ref{eq: DTW_similarity_methodology}).
Here, $K$ is the total number of comparisons arising from all pairs of light curves of $Z$ and J1023, i.e. $K = \sum_{j,i} \big( N_w(Z_j) + N_w(X_i) \big)$, where $Z_j$ and $X_i$ index each of the light curves of $Z$ and J1023, respectively. 
Note that although the DTW comparisons are performed separately for each prototype light curve and, when applicable, for the light curves of the system $Z$, all resulting DTW similarity values are pooled together to form a single ECDF, $\widehat{F}_Z(s)$.

The first application of this method is to characterize the intrinsic variability of the prototype source itself.
To do this, we define a functional ensemble with the ECDFs obtained from all prototype-prototype comparisons $\{\widehat{F}_{X_i X_j}(s)\}_{i < j}$.
The top left-panel in Fig.~\ref{fig: ECDFs_comparison} shows these ECDFs as grey shaded curves.
The median ECDF ($G_f(s)$; dotted red curve) appears as a central reference.
While this ensemble represents the whole range of heterogeneity expected from self-comparisons of the prototype system, not all ECDFs contribute equally to this variability.
In particular, some ECDFs show pairwise comparisons where the similarity distributions differ greatly from the ensemble's typical behavior, represented by the median $G_f(s)$, artificially broadening the functional range and making the prototype more difficult to characterize accurately.

To identify these ECDFs we quantify the deviation of each one from $G_f(s)$ considering the scalar discrepancy statistic $W_1$ defined as (see \cite{Vallender_1974}):
\begin{equation}
  D_{ij} \coloneqq W_{1}(\widehat{F}_{X_i X_j}, G_f) =
  \int_{-\infty}^{+\infty}
\big|
\widehat{F}_{X_i X_j} - G_f(s)
\big| \, \mathrm{d}s ~, 
\label{eq: Dij_distances}
\end{equation}
where $W_1$ denotes the first-order Wasserstein distance \citep[see, e.g,][]{wasserstein1969markov, dobrushin1970prescribing} between the $\widehat{F}_{X_i X_j}$ and $G_f(s)$. 
This corresponds to the area between the two ECDFs, providing a global measure of their discrepancy.
Large values of $D_{ij}$ indicate poorer agreement of the particular light curve with the median behavior of the system.
No parametric model, generative assumption, or hypothesis-testing framework is assumed. 

The set of distances $\{D_{ij}\}$ computed in Eq.~(\ref{eq: Dij_distances}) provides a quantitative distribution for how typical it is to find the prototype in a given observation behaving as it does on average. 
We identify outliers in this distribution using a criterion based on the median absolute deviation (MAD).  
Denoting by $\mathrm{med}(D)$ and 
$\mathrm{MAD}(D) = \mathrm{med}(|D_{ij} - \mathrm{med}(D)|)$ the median and MAD of $\{D_{ij}\}$, respectively, we define outliers ECDFs as those satisfying:
\begin{equation}
D_{ij} > \mathrm{med}(D) + k \times \mathrm{MAD}(D)~,
\label{eq: Dij_MAD}
\end{equation}
with $k=1$ chosen to isolate the core of the ensemble.

Let $N_{\mathrm{prot, f}}$ denote the number of ECDFs in the full functional ensemble. This number is 703, and results from $N_{\mathrm{prot, f}} = N_{X_i}\times (N_{X_i} - 1) / 2$, where $N_{X_i}=38$ is the number of prototype light curves used to construct the ensemble, see Table~\ref{tab: sample_summary}. 
After applying Eq.~(\ref{eq: Dij_distances}) for each one of 703 ECDFs, and filter them via Eq.~(\ref{eq: Dij_MAD}), the number of ECDFs composing the resulting functional ensemble is $N_{\mathrm{prot}}=$496.
The top right-panel of Fig.~\ref{fig: ECDFs_comparison} shows these ECDFs. 
The median ECDF ($G(s)$; solid black curve) appears as a reference curve.
It can be seen that $G(s)$ is indistinguishable from the median obtained from the reduced set of light curves selected to characterize the typical J1023 behavior.

\subsection{Confidence bands for DTW similarity }
\label{sec: confidence_bands}

\begin{figure*}
  \centering
  \includegraphics[width=\textwidth]{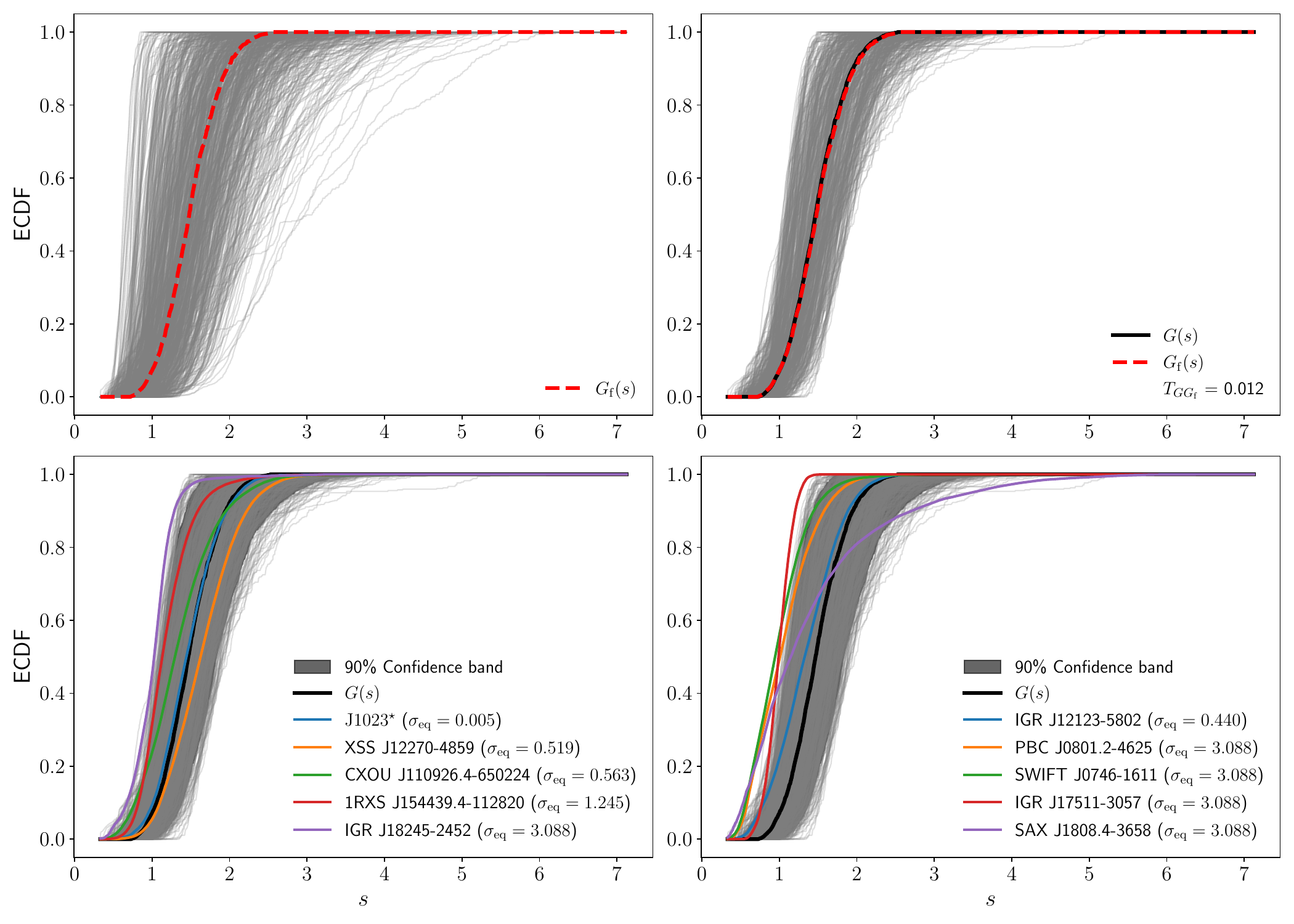}
\caption{Top left: All ECDFs coming from J1023's light curves self-comparisons. Their median, $G_{\mathrm{f}}(s)$, is displayed as a dotted red curve. 
Top right: The ECDFs filtered using Eq.~(\ref{eq: Dij_MAD}) are shown. The central tendency of the set of ECDFs from the left panel, $G_{\mathrm{f}}(s)$, is shown as a dotted red curve, and the one of the filtered ECDFs, $G(s)$, as a solid black curve. They are indistinguishable at this scale. 
The discrepancy between the medians, $T_{GG_{\mathrm{f}}}$, computed using Eq.~(\ref{eq: wasserstein_TZ}), is small, consistent with the visual match.
Bottom: ECDFs for the comparison of different systems, $Z$, with the prototype of the class, J1023. 
The left (right) panel corresponds to the candidate (control) systems.
The dark-shaded region represents the 90\% confidence band summarizing the heterogeneity among the prototype self-comparisons considering Eq.~(\ref{eq: Dij_MAD}).
The colored ECDFs are constructed from the pooled set of DTW-derived similarity values obtained by independently comparing the light curves of $Z$ with each available prototype light curve of J1023 denoted by $\widehat{F}_{Z}(s)$.
The solid black curve is the median of the prototype-prototype ECDFs, $G(s)$. 
}
  \label{fig: ECDFs_comparison}
\end{figure*}

From the ensemble of ECDFs representing the typical behavior of the prototype as computed in \S\ref{sec: similarity_ECDFs} a non-parametric pointwise $(1-\alpha)$ confidence band can be constructed at each fixed $s$ value as the region $[L_{\alpha}(s),\,U_{\alpha}(s)] = [\operatorname{quantile}_{\alpha/2}
        \left\{
        \widehat{F}_{X_iX_j}(s)
        \right\},\operatorname{quantile}_{1-\alpha/2}
        \left\{
        \widehat{F}_{X_iX_j}(s)
        \right\}
        ]$. 
It provides a reference against which to compare the ECDF $\widehat{F}_{Z}(s)$ of any source $Z$, indicating whether the similarity distribution of system $Z$ aligns with the intrinsic variability observed within the prototype. 
In this work, we adopt $\alpha = 0.1$, corresponding to a 90\% confidence band, shown as the dark-grey shaded region in Fig.~\ref{fig: ECDFs_comparison}.

\subsection{Quantifying similarity in a Gaussian-equivalent compatibility scale}
\label{sec: gaussian_equiv}
%
The departure of any system from the prototype-like behaviour can also be assessed using the scalar discrepancy statistics that measures deviations from $G(s)$ (see \S\ref{sec: confidence_bands}):

\begin{equation}
T_{Z} \coloneqq 
W_1\!\left( \widehat{F}_Z,\, G \right)
=
\int_{-\infty}^{+\infty}
\big|
\widehat{F}_Z(s) - G(s)
\big| \, \mathrm{d}s ~.
\label{eq: wasserstein_TZ}
\end{equation}
Large values of $T_{Z}$ indicate poorer agreement with the prototype class.
To interpret $T_{Z}$ probabilistically, these values will be calibrated against the range of deviations exhibited by the non-outlier prototype self-comparisons themselves; i.e., against the typical heterogeneity of J1023.
For the set of ECDFs selected in \S\ref{sec: similarity_ECDFs} we compute:

\begin{equation}
T_{X_iX_j} \coloneqq 
W_1\!\left( \widehat{F}_{X_iX_j},\, G(s) \right)
\qquad (i < j)~.
\label{eq: wasserstein_Tij}
\end{equation}
The collection $\{T_{X_iX_j}\}_{i < j}$, with a length of $N_{\mathrm{prot}}$, provides a finite reference distribution against which the deviation $T_Z$ can be interpreted.
For a given system $Z$, we define a reference tail value as:
\begin{equation}
r_Z =
\frac{
\#\{T_{X_iX_j} \ge T_{Z}\} + 1
}{
N_{\mathrm{prot}}  + 1}~,
\label{eq: reference_tail}
\end{equation}
which quantifies the relative position of $T_Z$ within the reference distribution.
In Eq.~(\ref{eq: reference_tail}), the numerator counts how many of the light curves of the prototype produce a discrepancy as large as the one observed for $Z$. 
This $r_Z$-value may then be converted into a Gaussian-equivalent compatibility scale via:
\begin{equation}
\sigma_{\mathrm{eq}}
=
\Phi^{-1}\!\left(1 - \frac{r_Z}{2}\right)~,
\label{eq: sigma_eq}
\end{equation}
where $\Phi^{-1}$ denotes the inverse cumulative distribution function of the standard normal distribution.
Equation~(\ref{eq: sigma_eq}) is purely monotonic and serves only to express relative extremeness on familiar statistical units.
The quantity $\sigma_{\mathrm{eq}}$ therefore provides a continuous scale of the compatibility between system $Z$ and the prototype behaviour.
It complements the confidence-band analysis by providing a scalar summary of how the similarity distribution of systems relates to the intrinsic variability shown by the prototype source.
Lower values indicate greater morphological compatibility, whereas higher values indicate increasingly atypical morphology relative to the prototype.
These values should be understood strictly as measures of compatibility based on a reference, not as mathematical probabilities, hypothesis tests, or rejection criteria.
In particular, a value such as $\sigma_{\rm eq}=3$ should not be interpreted as a conventional $3\sigma$ rejection threshold.
The finite extent of the $\sigma_{\mathrm{eq}}$ scale reflects the finite size of the reference distribution and sets the maximum discriminating power achievable with the available prototype observations.
For a given system $Z$, the quantity $\#\{T_{X_iX_j}\geq T_Z\}$ in Eq.~(\ref{eq: reference_tail}) can only take integer values, and therefore the resolution of $r_Z$ scales as $1/N_{\rm prot}$. 
In our case, where $N_{\rm prot}=496$ (see \S\ref{sec: similarity_ECDFs}), the maximum value, $\sigma_{\rm eq}=3.088$, is obtained when $\#\{T_{X_iX_j}\geq T_Z\}=0$. 
This corresponds to the case in which all prototype self-comparisons are closer to the median $G(s)$ than the system $Z$ itself.
The next possible value occurs for $\#\{T_{X_iX_j}\geq T_Z\}=1$, yielding $\sigma_{\rm eq}=2.876$. 
In this case, at least one ECDF within the representative prototype ensemble is as discrepant from $G(s)$ as, or more discrepant than, the system under consideration. 
More generally, increasing values of $\#\{T_{X_iX_j}\geq T_Z\}$ indicate that a larger fraction of the prototype self-comparisons show deviations from $G(s)$ at least as large as that of the system $Z$.
Values approaching the upper end of the observed range therefore indicate morphology more discrepant than nearly all prototype self-comparisons, rather than a formal statistical rejection of prototype-like behaviour.

\section{Results}
\label{sec: results}

%
The moving-window process was carried out with $w_s = 40$ bins and $s_s = 1$ bin, ensuring maximum overlap between consecutive segments, and the largest possible number of windows. 
The DTW alignment was computed using a Sakoe-Chiba band with $w_w = 4$ bins.
Throughout the analysis, the hyperparameters introduced in \S~\ref{sec: basics_dtw} and \S~\ref{sec: methodology_dtw_tmsp} were held constant, see  Appendix \ref{appendix: hyp}.
Figure~\ref{fig: ECDFs_comparison} shows the results obtained. 

Among the lowest values of $\sigma_{\mathrm{eq}}$, and in particular, in the second lowest, we find XSS J12270-4859, CXOU J110926.4-650224, 1RXS J154439.4-112820; the former is a confirmed tMSP, while the latter two are well-established candidates. 
A known tMSP, IGR J18245-2452 appears with a high score. 
This is the result of it having a less structured light curve (showing a bin-by-bin step-like appearance), which, even qualitatively, does not resemble those of the prototype.
Because the source is in a low-count regime, each 30~s time bin contains only a small integer number of detected photons. As a result, the background-subtracted count rate takes on discrete values rather than varying smoothly, producing a step-like appearance.
Consequently, applying DTW to this Chandra light curve is less straightforward than to the smoother XMM-Newton data, and the results should be interpreted with caution.
We retain the comparison with IGR J18245-2452 here because it is a bona fide tMSP, while noticing that, a priori, we do not expect a close behavioral matching.
See Appendix \ref{appendix: src_sample} for references and details of the light curves used.

One source, IGR J12123-5802, a nova-like CV appears with a low $\sigma_{\mathrm{eq}}$, providing an interesting handle regarding what the methodology is sensitive to (see Appendix \ref{appendix: preprocessing_light_curves} and especially Fig. \ref{fig: others-light-curves} for the system's light curve).
Generic accretion-driven flickering can mimic switching on short timescales. 
In CVs, variable absorption from structures in the disk, stream, or a disc rim can produce repeated dips or intervals of lower count rate. When binned at 30 s and viewed in segments, this can resemble alternating high/low intervals.
Also, as we keep for each window only the best DTW match in the other light curve, a source can score well if it contains a non-trivial subset of windows that resemble some J1023 windows, even if the global behaviour is different.
Thus, there is no contradiction in a nova-like CV achieving moderate similarity to a tMSP prototype under a morphology-only metric with best-window pooling. 
Rather it is a warning that the method can also capture generic accretion variability patterns that are not unique to tMSPs.
In fact, two of the prototypical tMSPs, PSR~J1023+0038 and XSS~J12270--4859, were initially classified as CVs \citep{Bond2002,deMartino2010}.
%

\section{Concluding remarks}

We introduce a quantitative, morphology-based framework to compare X-ray light curves of transitional millisecond pulsar candidates against a prototype source, using DTW and ECDFs to summarize the resulting similarity distributions. 
The aim is to replace purely visual classification with a reproducible and calibrated comparison scheme. Our conclusions are:

\begin{itemize}
\item The intrinsic heterogeneity of the prototype, PSR J1023+0038, can be characterized through the ensemble of prototype-prototype ECDFs, whose median $G(s)$ provides a natural central reference and whose spread defines the expected level of variability compatible with the class.

\item The methodology succeeds in separating, at the ensemble level, confirmed or candidate tMSPs from the control sample in a statistically interpretable way, even though the classification is based only on light-curve morphology and does not use source labels or physical priors.

\item The only clear exception is that of a known tMSP, IGR J18245-2452, which appears with a high score. This was expected, as the light curve used shows a bin-by-bin step-like appearance.

\item Restricting the prototype band to the ECDFs closest to the median makes the compatibility criterion more stringent, but does not substantially shift the median reference curve itself. 
The resulting filtered ensemble therefore characterizes the representative, internally consistent behaviour of the prototype against which the compatibility of other systems is assessed. 
Thus, the main effect of narrowing the heterogeneity band is a stricter statistical calibration, not a qualitative redefinition of the typical prototype behaviour.

\item The method is sensitive to morphology, not directly to source class. Accordingly, a non-tMSP system can achieve moderate similarity to the prototype if it contains a non-negligible subset of light curve segments that resemble some of the J1023 windows, even when its overall behaviour is different.
In particular, different accretion-driven states in systems that are not tMSP candidates can individually produce ECDFs that lie close to the prototype median. This does not imply that such systems are physically akin to tMSPs as a whole; rather, it shows that some short-timescale accretion patterns are not unique to transitional pulsars and can mimic isolated pieces of their phenomenology under a morphology-only metric.

\item For that reason, the most meaningful discriminator is not the existence of a single ECDF close to $G(s)$, but the aggregate behaviour of the full comparison ensemble associated with a source, as represented by $F_Z$ or, similarly, by the median of that ensemble.

\item As a particular case, IGR J12123-5802 is informative rather than contradictory: it highlights that the method can capture generic accretion variability patterns that overlap morphologically with parts of the tMSP phenomenology, and thus clarifies both the scope and the present limitation of the approach.

\end{itemize}

Overall, the analysis supports the use of DTW-based ECDF comparisons as a useful quantitative screening tool for tMSP-like behaviour, while also showing that morphology alone is not fully specific. 
The method is informative even in cases when it does not uniquely identify class membership: it quantifies how far morphology alone can take the inference, and therefore where additional diagnostics become necessary.

The method is therefore best interpreted as a first-stage similarity filter, certainly beyond a visual comparison.
A practical use of the method is candidate triage: when several sources show qualitatively suggestive variability, the DTW-based score provides a reproducible way to prioritize those most compatible with established tMSP phenomenology for deeper observational campaigns.
Because the procedure is automatic once light curves are prepared, it can be applied systematically to archival datasets to search for previously unrecognized tMSP-like variability patterns.
While here used for assessing individual candidates to tMSPs, it does not preclude additional physical or observational diagnostics. 
It may play a role in automatic catalog searches for candidates presenting a particular light curve behavior, beyond that of tMSPs.

\section*{Acknowledgements}

This work has been supported by the grant PID2024-155316NB-I00 funded by MICIU /AEI /10.13039/501100011033 / FEDER, UE and CSIC PIE 202350E189. This work was also supported by the Spanish program Unidad de Excelencia María de Maeztu financed by MCIN/AEI/10.13039/501100011033, and by the MaX-CSIC Excellence Award MaX4-SOMMA-ICE. 
F.C.Z. is supported by a Ramon y Cajal fellowship (grant agreement RYC2021-030888-I) and by the Spanish grant ID2023-153099NA-I00.
GI is supported by a Juan de La Cierva fellowship (JDC2024-053550-I).
A.P., D.dM. acknowledges support by INAF (Research Grant FANS and PULSE-X, PI: Papitto), the Italian Ministry of University and Research (PRIN MUR 2020, Grant 2020BRP57Z, GEMS, PI: Astone), and Fondazione Cariplo/Cassa Depositi e Prestiti (Grant 2023-2560, PI: Papitto). 
M.C.B. acknowledges support from the INAF Fundamental Research Grant XBOOM.
We thank two anonymous referees for constructive interactions.

\section*{Data availability}
The data underlying this article will be shared on reasonable request to the corresponding author.

\section*{Conflicts of Interest}
Authors declare no conflict of interest.

  \bibliographystyle{mnras}
\bibliography{biblio}

\begin{appendix}

\section{Source sample \& observational details}
\label{appendix: src_sample}

This appendix summarizes the nature of the sources included in our comparison sample and the observational material used for each of them. The emphasis here is put on the literature most relevant to the accreting or sub-luminous states from which the light curves analyzed in this work were extracted.

\begin{itemize}

    \item \textbf{PSR J1023+0038 (J1023).} J1023 is the prototype and best-studied confirmed tMSP. It was first recognized as an unusual CV candidate, then identified as a radio MSP with evidence of a previous accretion disc, and it re-entered an accretion-dominated sub-luminous state in 2013, remaining the benchmark source for tMSP mode-switching phenomenology \citep{Bond2002,Archibald2009,Stappers2014J1023,Patruno2014, Bogdanov2015J1023,Archibald2015J1023}. In this work we use 17 XMM-Newton/EPIC observations obtained during this current active state, since they define the prototype ensemble against which the rest of the sample is compared. These observations encompass the canonical rapid switches among high, low, and flaring modes. Later campaigns show that the same phenomenology persists across the subsequent monitoring of the source \citep{Campana2016J1023,CotiZelati2018J1023,Papitto2019J1023,MiravalZanon2022J1023,Baglio2023J1023}.

    \item \textbf{XSS J12270-4859 / PSR J1227-4853 (J12270).} XSS J12270-4859 is a confirmed tMSP that spent years in a sub-luminous disc state before transitioning in late 2012 to a radio-pulsar state \citep{deMartino2010,deMartino2013,Bassa2014J12270,Roy2015J12270}. The source is especially relevant here because its pre-transition X-ray behaviour closely matches the tMSP phenomenology defined by J1023, and coherent X-ray pulsations were later detected in the 2011 sub-luminous-state data \citep{Papitto2015J12270}. We consider two observations overall; the two key disc-state XMM-Newton observations are those performed in 2009 and 2011. For these, we use the summed EPIC- MOS1+MOS2 light curves.  

    \item \textbf{IGR J18245-2452 / PSR J1824-2452I (J1824, M28I).} IGR J18245-2452 is the first system directly observed to swing between accretion-powered and rotation-powered states, providing the clearest proof of the recycling link between LMXBs and radio MSPs \citep{Papitto2013,Linares2014J18245,Ferrigno2014J18245}. Unlike J1023 and J12270, its transitional behaviour was caught during a bright X-ray outburst in the globular cluster M28 in 2013, including extreme ``hiccup'' accretion and order-of-magnitude intensity changes on short time-scales \citep{Ferrigno2014J18245}. In our sample, we consider a Chandra light curve in which the source is observed in a subluminous state.

    \item \textbf{1RXS J154439.4-112820 / 3FGL J1544.6-1125 (J1544).} J1544 is one of the strongest tMSP candidates identified among previously unassociated Fermi-LAT sources \citep{BogdanovHalpern2015J1544,Bogdanov2016J1544}. Its optical spectrum shows an accretion disc, while the X-ray light curve displays the same rapid high/low-mode switching that characterizes confirmed tMSPs \citep{BogdanovHalpern2015J1544,Bogdanov2016J1544}. Recent simultaneous radio/X-ray observations further strengthen the tMSP interpretation by showing behaviour consistent with the radio/X-ray phenomenology already seen in J1023 \citep{Gusinskaia2025J1544}. A recent dedicated multi-band study by \citet{Illiano2025J1544} further probed the source's mode switching and broad-band variability, reinforcing its standing as one of the best current tMSP candidates. In this work we consider three XMM-Newton observations.

    \item \textbf{CXOU J110926.4-650224 (J1109).} J1109 is a strong tMSP candidate that has remained in a prolonged sub-luminous disc state since its recognition as the X-ray counterpart of an INTEGRAL hard-X-ray source \citep{CotiZelati2019J1109,CotiZelati2021J1109}. Multiwavelength follow-up has shown the rapid bimodal variability, broad-band spectral properties, and variable radio emission expected for the class, while the latest Fermi-LAT analysis now supports its association with a GeV counterpart \citep{CotiZelati2024J1109,Manca2025J1109}. We consider two XMM-Newton observations for this source.

\end{itemize}


\begin{table*}
\centering
\caption{Observational log for the sources used in this work. Columns list the source name, instrument, observation identifier, observation date, and reference. Light curves are binned with 30~s steps}.
\label{tab: obslog}
\footnotesize
\resizebox{\textwidth}{!}{
\begin{tabular}{llllcc}
\toprule
Source & Instrument & Obs.\ ID & Date & Duration (ks) & Reference \\
\midrule
J1023 & XMM-Newton/EPIC & 0720030101 & 2013-11-10 & 138.1 & \citep{Bogdanov2015J1023,Archibald2015J1023} \\
J1023 & XMM-Newton/EPIC & 0742610101 & 2014-06-10 & 131.1 & \citep{Bogdanov2015J1023,Archibald2015J1023} \\
J1023 & XMM-Newton/EPIC & 0748390101 & 2014-11-21 & 35.7 & \citep{Campana2016J1023,Jaodand2016J1023} \\
J1023 & XMM-Newton/EPIC & 0748390501 & 2014-11-23 & 36.2 & \citep{Campana2016J1023,Jaodand2016J1023} \\
J1023 & XMM-Newton/EPIC & 0748390601 & 2014-11-28 & 22.0 & \citep{Campana2016J1023,Jaodand2016J1023} \\
J1023 & XMM-Newton/EPIC & 0748390701 & 2014-12-17 & 35.8 & \citep{Campana2016J1023,Jaodand2016J1023} \\
J1023 & XMM-Newton/EPIC & 0770581001 & 2015-11-11 & 32.4 & \citep{Campana2016J1023,Jaodand2016J1023} \\
J1023 & XMM-Newton/EPIC & 0770581101 & 2015-11-13 & 24.0 & \citep{Campana2016J1023,Jaodand2016J1023} \\
J1023 & XMM-Newton/EPIC & 0783330301 & 2015-12-09 & 27.7 & \citep{Campana2016J1023,Jaodand2016J1023} \\
J1023 & XMM-Newton/EPIC & 0784700201 & 2016-05-08 & 127.9 & \citep{CotiZelati2018J1023} \\
J1023 & XMM-Newton/EPIC & 0794580801 & 2017-05-23 & 26.0 & \citep{Papitto2019J1023} \\
J1023 & XMM-Newton/EPIC & 0794580901 & 2017-05-24 & 24.5 & \citep{Papitto2019J1023} \\
J1023 & XMM-Newton/EPIC & 0803620201 & 2017-05-08 & 25.3 & -- \\
J1023 & XMM-Newton/EPIC & 0803620301 & 2017-05-10 & 29.0 & -- \\
J1023 & XMM-Newton/EPIC & 0803620401 & 2017-05-16 & 24.0 & -- \\
J1023 & XMM-Newton/EPIC & 0803620501 & 2017-06-13 & 24.0 & \citep{MiravalZanon2022J1023} \\
J1023 & XMM-Newton/EPIC & 0864010101 & 2021-06-03 & 64.5 & \citep{Baglio2023J1023} \\
J1109 & XMM-Newton/EPIC & 0764344301 & 2018-06-20 & 68.0 & \citep{CotiZelati2019J1109} \\
J1109 & XMM-Newton/EPIC & 0864190201 & 2022-03-02 & 24.0 & \citep{CotiZelati2024J1109} \\
J1544 &  XMM-Newton/EPIC-pn & 0800280101 & 2018-01-28 & 133 & \citep{Illiano2025J1544} \\ 
J1544 &  XMM-Newton/EPIC-pn & 0923466201 & 2024-02-08 & 60 & \citep{Illiano2025J1544} \\ 
J1544 &  XMM-Newton/EPIC & 0724080101 & 2014-02-16 & 42.2 & \citep{BogdanovHalpern2015J1544} \\
J12270 & XMM-Newton/EPIC-MOS & 0551430401 & 2009-01-05 & 31.49 & \citep{deMartino2010} \\
J12270 & XMM-Newton/EPIC-MOS & 0656780901 & 2011-01-01 & 31.02 & \citep{deMartino2013} \\ 
J1824 & Chandra/ACIS-S & 9132 & 2008-08-07 & 142.26 & \citep{Papitto2013, Linares2014J18245} \\
PBC J0801.2-4625 & XMM-Newton/EPIC-pn & 761120301 & 2015-12-04 & 30 & \citep{Bernardini_2017} \\
SWIFT J0746-1611 & XMM-Newton/EPIC-pn & 761120401 & 2016-04-01 & 34.5 & \citep{Bernardini_2017} \\
IGR J12123-5802 & XMM-Newton/EPIC-pn & 671850601 & 2012-01-07 & 38.2 & \citep{Bernardini2013} \\
IGR J17511-3057 & XMM-Newton/EPIC & 0954191401 & 2025-03-03 & 20 & \citep{Illiano2026_J17511} \\
SAX J1808.4-3658 & XMM-Newton/EPIC & 0884700801 & 2022-09-09 & 113.3 & \citep{Ballocco_SAX1808} \\
\bottomrule
\end{tabular}
}
\end{table*}

Regarding the sources that are not tMSPs or candidates, 

\begin{itemize}

    \item {\bf PBC J0801.2-4625.} It is a magnetic CV of the Intermediate Polar type with a spin period of 22min \citep[see][]{Bernardini_2017, Halpern2018, Irving2024}. 

    \item {\bf IGR J12123-5802.} It is supposed to be a magnetic CV, but no fast periodicity is found. It is classified as a nova-like CV \citep[see][]{Bernardini2013}. 

    \item {\bf SWIFT J0746-1611.} It is a magnetic CV of the IP type that, when in a lower state, did not show fast periodicity and was suggested to be possibly a LMXB. However, when found a few years later in a high state, a spin of 38min was found (see again \citealt{Bernardini_2017} and \citealt{Bernardini2019}). For this source, the light curve is in a low state, which could be helpful to compare with the tMSP light curves. 

    \item {\bf IGR J17511-3057.} 
     The EPIC light curve, binned at 30 s, from the XMM-Newton observation of the AMSP was obtained during the late stages of its recent outburst \citep{Illiano2026_J17511}. At that point, the X-ray luminosity had decreased below 7e33 erg/s, consistent with levels observed in tMSPs in the sub-luminous state. Even at this level, we would expect to be in the propeller regime, but we detected coherent X-ray pulsations. This behavior was also observed in another AMSP using NICER \citep{Bult_2019}.

    \item {\bf SAX J1808.4-3658.} Light curve of the AMSP during the final stage of its 2022 outburst, observed with EPIC pn and binned at 30 s (see \citealt{Ballocco_SAX1808}). The observation was performed during the so-called reflaring state.

\end{itemize}

All XMM-Newton observations listed in Table~\ref{tab: obslog} were processed following standard procedures with the XMM-Newton Science Analysis System (SAS; \citealt{Gabriel2004SAS}). After screening out intervals affected by enhanced particle background, source products were extracted from circular regions centered on the target for imaging-mode observations, with background regions selected away from the source but on the same CCD. For timing-mode observations, source and background events were instead extracted from strip-shaped regions centered on, and offset from, the source columns, respectively. Background-subtracted light curves in the 0.3--10~keV band were then produced with \texttt{epiclccorr}, which applies the standard exposure and instrumental corrections. Whenever more than one EPIC detector contributed simultaneous good-time intervals, the corresponding light curves were combined over the strictly overlapping intervals. The final light curves were binned at 30~s, which provides a practical compromise between signal-to-noise in individual bins and sensitivity to short low-mode intervals.

The Chandra data for J1824 were reprocessed with \texttt{chandra\_repro} in CIAO~4.18 using CALDB~4.12.3 \citep{Fruscione2006}. We extracted source events from a circular aperture of radius $1.5''$ centered on the source position. The background was estimated from an annulus with inner and outer radii of $6''$ and $16''$, after masking neighboring point sources detected in the same field. Background-subtracted light curves were then produced in the 0.5--8.0~keV band with \texttt{dmextract}, which applies the standard exposure correction on a bin-by-bin basis, using 30~s time bins.

\section{Preprocessing light curves}
\label{appendix: preprocessing_light_curves}

The preparation of the light curves for application of the developed methodology consists of three main steps: treatment of negative count rate values, identification of temporal discontinuities, and removal of flaring modes.
Negative count rate values are set to zero, under the assumption that they arise from statistical fluctuations around a null signal and do not represent physical emission.
Temporal discontinuities are identified as gaps in which no count value is recorded. 
Since the light curves are binned consecutively in time with a resolution of 30~s, any time difference between two adjacent bins exceeding 30~s (corresponding to 1 bin) is interpreted as a temporal jump.
In such cases, the light curve is split into independent segments (resulting in different light curves). 
Only segments with a minimum temporal length of 3600~s (1 hr) are retained for further analysis.
The removal of flaring modes is carried out using the count-rate distribution, as explained in the main text.
Each resulting light curve is normalized using the min-max scaler:
\begin{eqnarray}
X^{\dag}=\frac{x_i-x_{\min}}{x_{\max}-x_{\min}}~,
\label{eq: min-max_scaler}
\end{eqnarray}
where $x_i$ denotes the flux value at $i$-bin-index, and $x_{\min}$ and $x_{\max}$ are the minimum and maximum flux values of the corresponding light curve.
Equation~(\ref{eq: min-max_scaler}) ensures that each light curve is evaluated mainly based on its morphology, regardless of flux variations, adjusting its values between 0 and 1.
Table~\ref{tab: sample_summary} shows the number of resulting light curves for each source. 
\begin{table}
\centering
\caption{Summary of the preprocessed sample. 
The number of observational campaigns used for each system is shown, along with the resulting number of light curves (LCs) obtained after the preprocessing steps described in \S\ref{sec: preprocessing_light_curves}. 
Systems classified in the literature as confirmed and candidate tMSPs are shown in boldface.
}
\label{tab: sample_summary}
\begin{tabular}{lcc}
\toprule
\textbf{System} & \textbf{Campaigns} & \textbf{Resulting LCs} \\
\midrule
\textbf{J1023}           & 17 & 38 \\
\textbf{CXOU J110926.4-650224} & 2  & 3  \\
\textbf{XSS J12270-4859} & 2  & 2  \\
\textbf{1RXS J154439.4-112820} & 3  & 9  \\
\textbf{IGR J18245-2452} & 1  & 1  \\
PBC J0801.2-4625 & 1  & 1  \\
SWIFT J0746-1611 & 1  & 1  \\
IGR J12123-5802  & 1  & 1  \\
IGR J17511-3057  & 1  & 1  \\
SAX J1808.4-3658 & 1  & 2  \\
\bottomrule
\end{tabular}
\end{table}

\begin{figure*}
  \centering
  \includegraphics[width=0.88\textwidth]{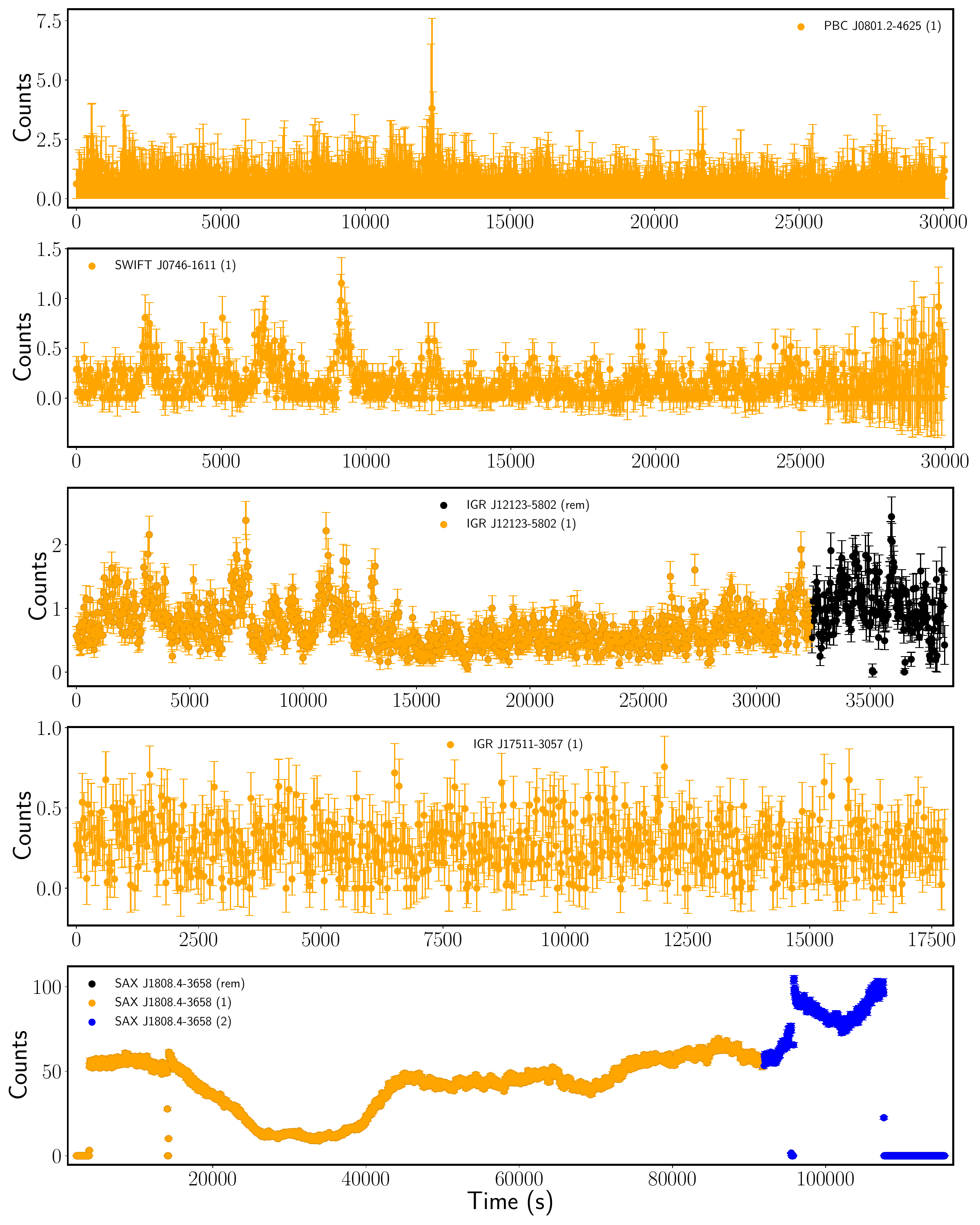}
\caption{Light curves for other sources, not classified as tMSPs or candidates.
}
  \label{fig: others-light-curves}
\end{figure*}

Figures~\ref{fig: proto-light-curves}, \ref{fig: proto-light-curves_continued}, and \ref{fig: proto-light-curves_continued_2} show the light curves for the prototype transitional pulsar source J1023.
Each panel represents a different observation.  
The removed bins (due to them pertaining to a flare according to the definition, or because of a short duration of consecutive bins) are shown in black (labelled as rem).
Different colors explicitly mark the light curve segments retained for further analysis. 
Figures~\ref{fig: candidate-light-curves}, \ref{fig: candidate-light-curves_continued}, and \ref{fig: others-light-curves} show the considered light curves for the transitional pulsar candidates and the other sources in the comparison, respectively.

\begin{figure*}
  \centering
  \includegraphics[width=0.88\textwidth]{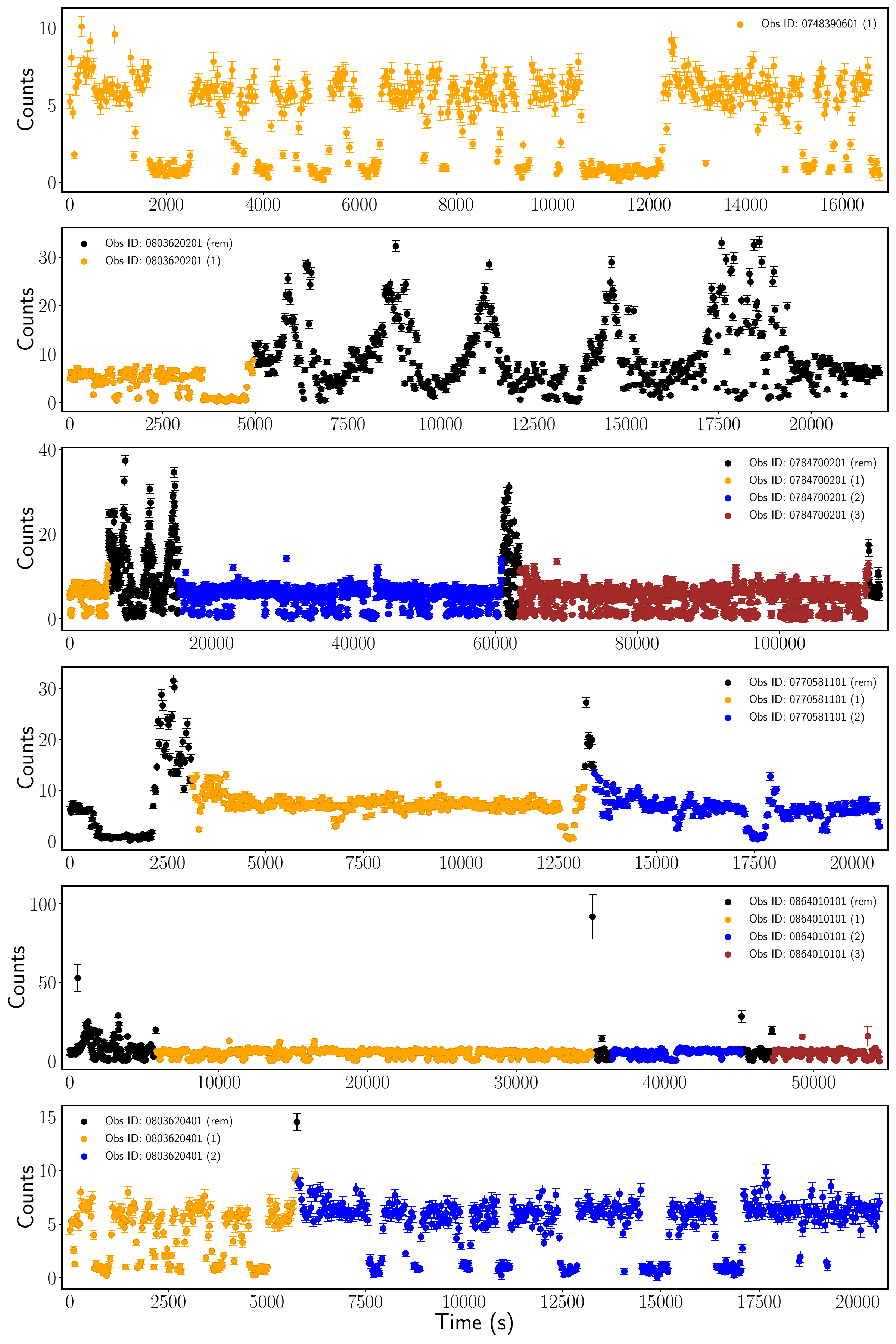}
\caption{Example of the light curve treatment using the prototype transitional pulsar source J1023. See text for details. 
}
  \label{fig: proto-light-curves}
\end{figure*}

\begin{figure*}
  \centering
  \includegraphics[width=0.88\textwidth]{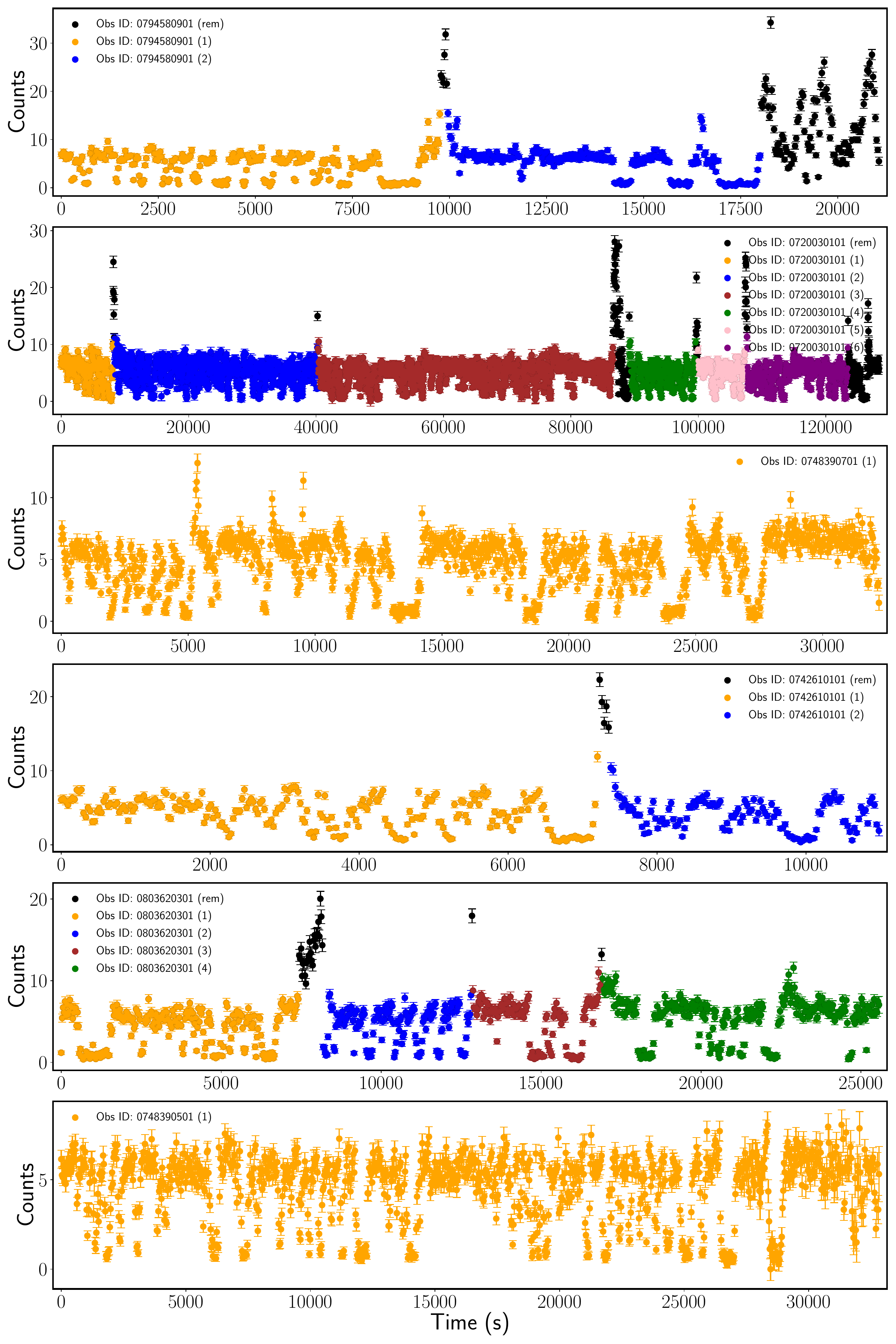}
\caption{Continued. Example of the light curve treatment using the prototype transitional pulsar source J1023. See text for details. 
}
  \label{fig: proto-light-curves_continued}
\end{figure*}

\begin{figure*}
  \centering
  \includegraphics[width=0.88\textwidth]{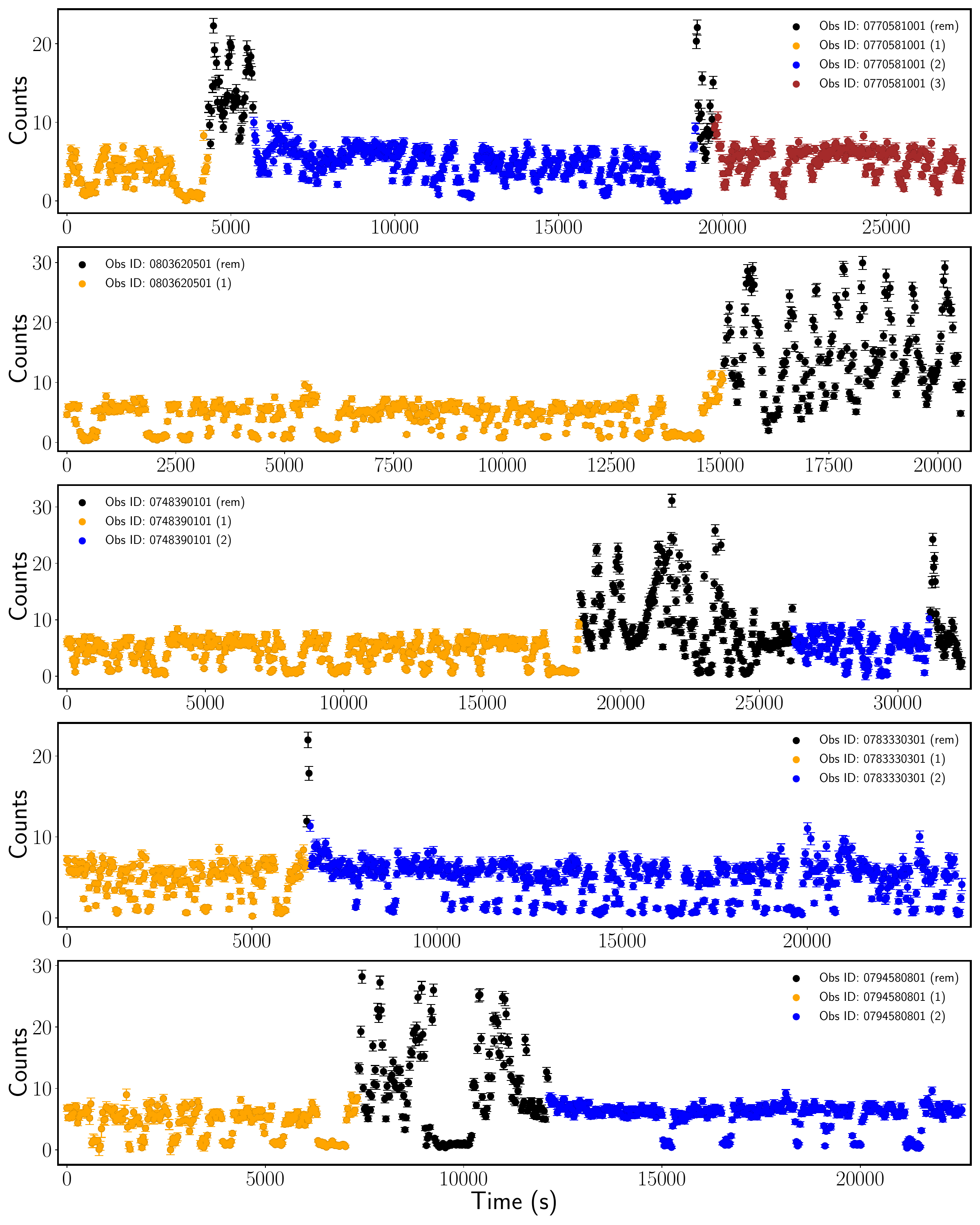}
\caption{Continued. Example of the light curve treatment using the prototype transitional pulsar source J1023. See text for details. 
}
  \label{fig: proto-light-curves_continued_2}
\end{figure*}

\begin{figure*}
  \centering
  \includegraphics[width=0.88\textwidth]{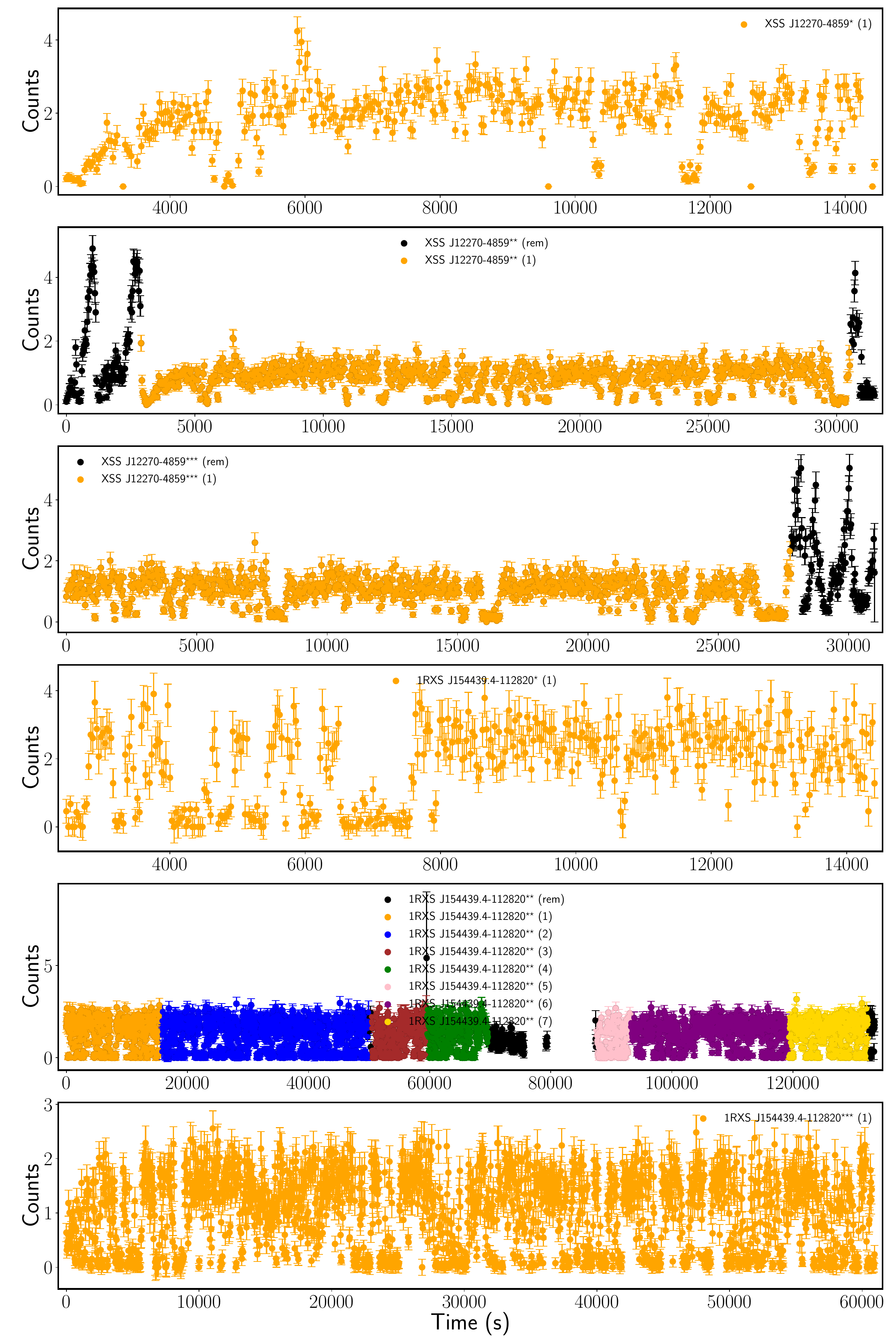}
\caption{Candidates light curves used in the comparison.
}
  \label{fig: candidate-light-curves}
\end{figure*}

\begin{figure*}
  \centering
  \includegraphics[width=0.88\textwidth]{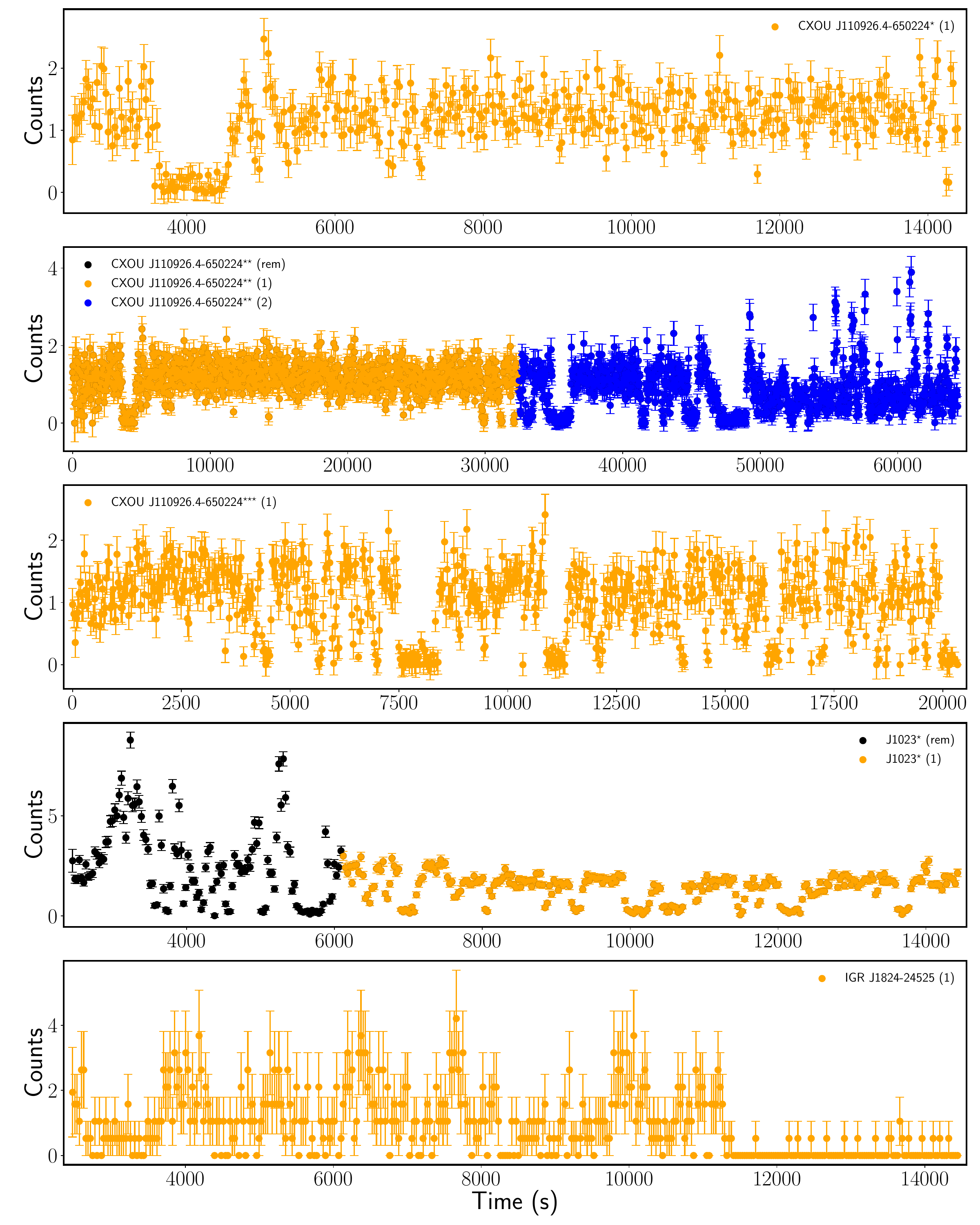}
\caption{Continued. Candidates light curves used in the comparison.
}
  \label{fig: candidate-light-curves_continued}
\end{figure*}

\section{Examples of moving window process}
\label{appendix: mov-win}
\begin{figure*}
  \centering
  \includegraphics[width=\textwidth]{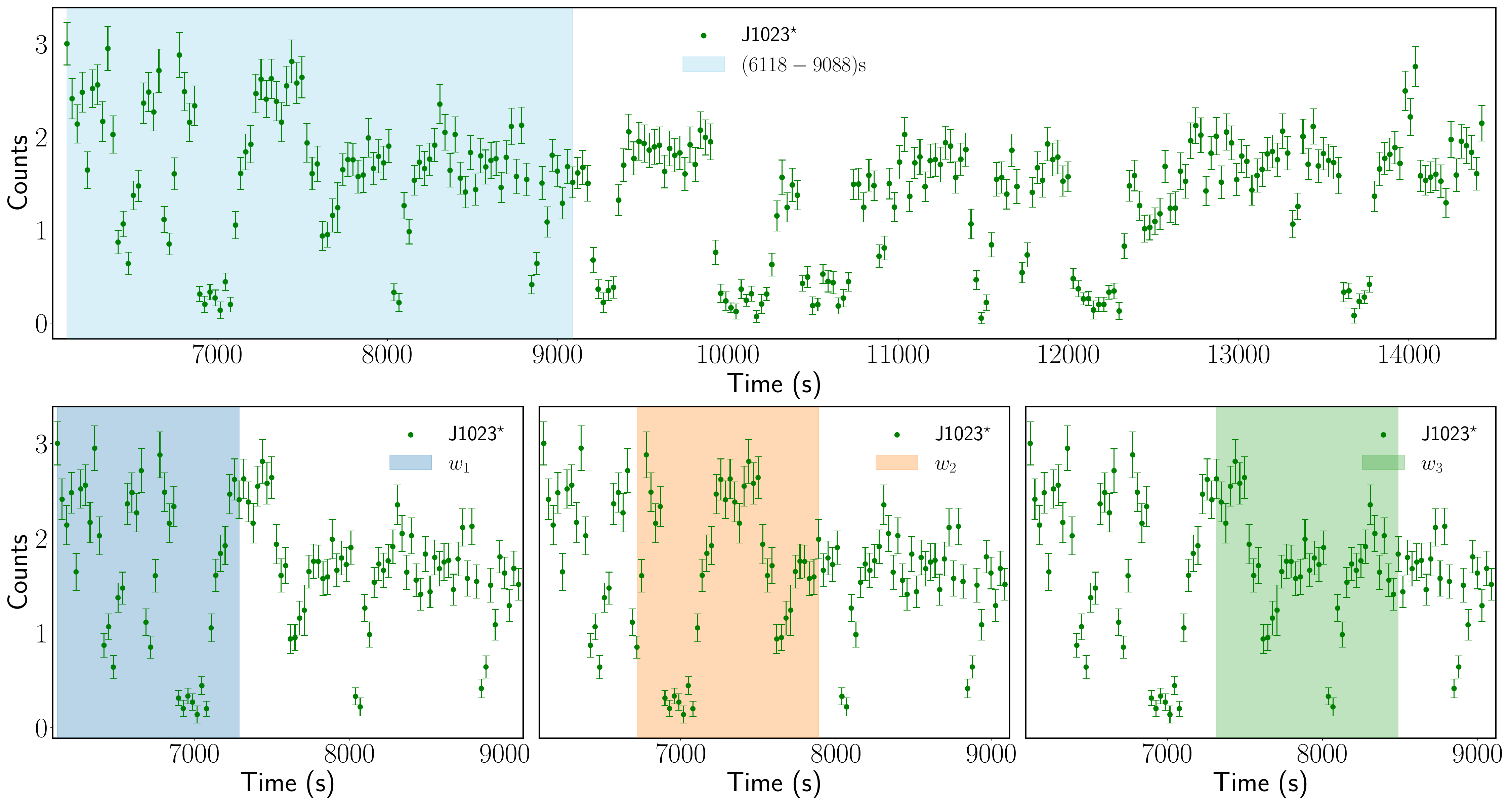}
\caption{First row: Excerpt of a light curve for 1RXS J154439.4-112820$^{\star}$, with the time interval 2472-5442~s indicating the range selected for detailed analysis using the moving-window process shown in the second row.
Second row: Zoom of the light blue region in the first row, with representative windows highlighted. Each panel shows a different window $w$ from the moving-window process, with $ w_s = 40$ bins and a step size of $s_s = 20$ bins. The shaded region indicates the temporal extent of the selected $w$ (see Eq.~ (\ref{eq: time_series_windows})). The $s_s$ is chosen to be sufficiently large to maximize visibility.
}
  \label{fig: moving_window}
\end{figure*}

Figure~\ref{fig: moving_window} shows an example of the moving window technique.
The process in Eq.~(\ref{eq: number_windows}) starts by setting the first window at the beginning of the light curve (i.e., the first bin index, or 0-bin-index; the additional $+1$ in Eq.~(\ref{eq: number_windows}) accounts for this), using $w_s$. 
The second window starts at the $s_s$-bin-index, the third at the $2\times s_s$-bin-index, and so on. 

\section{Alternative DTW constraints}
\label{appendix: alternative_dtw_constraints}

The Sakoe-Chiba band and the Itakura parallelogram impose different restrictions on the admissible DTW warping path.
However, it is useful to clarify two aspects of the present methodology and the motivation for adopting the Sakoe-Chiba constraint.
First, the moving-window process used in our work and the DTW warping constraint operate at different stages of the analysis (see Appendix~\ref{appendix: mov-win} for more details). 
The moving window defines the local light-curve segments that we compare, whereas the Sakoe-Chiba or Itakura constraints determine which alignments are allowed within each pair of segments. 
They are therefore not alternative approaches. 
In our implementation, 1200~s windows are shifted by one 30-s bin at a time.
Consequently, although an individual window may intersect the boundary of a high or low mode, structures shorter than the window duration are sampled repeatedly at different positions and, except near the boundaries of a light-curve segment, are fully contained in multiple windows. 
The method therefore does not rely on a single arbitrary segmentation of the mode-switching behavior. 
Second, the  Sakoe-Chiba constraint, for equal-length windows, imposes a uniform maximum temporal displacement around the diagonal of the accumulated cost matrix throughout the comparison \citep{Müller_2007}, while leaving the DTW algorithm free to select the minimum-cost path within the admissible region. 
In contrast, the Itakura parallelogram imposes a position-dependent admissible region and additional slope restrictions. 
Both approaches can limit pathological warping, but they do so differently.
We do not think the Itakura constraint is inherently more appropriate for the present physical problem. 
In particular, because the positions of our moving windows are not tied to physical mode boundaries, the same high-to-low or low-to-high transition may occur near the beginning, center, or end of different overlapping windows.
We have no physical reason to allow greater temporal deformation when a given feature lies near the center of a window than when the same feature occurs closer to its boundaries.
We also explicitly examined the degree of warping permitted by the Sakoe-Chiba constraint. 
As described in Appendix~\ref{appendix: hyp}, we explored $w_w =$ 1, 2, 4, 8, and 20, as well as the unconstrained case. 
Intermediate values produced stable similarity results, whereas larger values permitted increasingly high-multiplicity alignments in which individual bins could be matched to many bins in the other light curve. 
We adopted $w_w =4$ as a compromise that allows moderate temporal deformation while retaining physically interpretable alignments.
Thus, we do not apply the Sakoe-Chiba constraint by default; we select it after examining how its allowed warping range affects the results.

Existing alternatives include “dynamic windowing,” where the general idea is to use a DTW constraint that becomes more flexible around rapid transitions and more restrictive during stable intervals.
Adaptive DTW constraints of various forms have indeed been proposed in the literature (see, e.g., R-K Band in \cite{Ratanamahatana_Keogh}, sDTW in \cite{Candan_2012_sDTW}, or LDTW in \cite{Zhang_LDTW}). 
However, “dynamic windowing” does not, by itself, define a unique DTW prescription. 
Our purpose here is instead to compare the complete morphology of each local light-curve segment without assigning a priori greater weight or greater temporal flexibility to the transitions than to the high- and low-state intervals themselves. 
The durations of these states may also be part of the morphology we intend to retain, or at least modify as little as possible. 
A smaller DTW cost obtained by compressing a long plateau onto a much shorter one would therefore not necessarily represent a physically preferable match.
For these reasons, we retain the symmetric Sakoe-Chiba constraint in the present analysis.

\section{Model hyperparameters}
\label{appendix: hyp}

This methodology involves three user-defined hyperparameters, $w_s$, $s_s$, and $w_w$, which act at different levels of the analysis.

The window size $w_s$ defines the characteristic timescale over which the light curves are compared and constitutes the method's primary scale. 
Its choice is constrained by the total length of the light curves, as it determines both the temporal resolution and the number of segments (windows) that can be extracted according to Eq.~(\ref{eq: number_windows}), together with $s_s$. 
Increasing $w_s$ reduces the number of available windows while emphasizing larger-scale structures.

Given the minimum light curve duration of 3600~s (see \S~\ref{sec: preprocessing_light_curves}), we set $w_s = 40$ bins, equivalent to 1200~s, corresponding to one third of the minimum allowed light curve length.
This choice balances the need to capture global features while retaining a sufficient number of windows for statistical characterization.
The step size $s_s$ determines how these windows are sampled. 
In this work, we adopt $s_s = 1$.  
This choice includes all possible window positions and maximizes the sampling of the light curve.
This leads to strongly overlapping windows.
In contrast, setting $s_s = w_s$ suppresses the moving window process.
Once the windows are defined, the parameter $w_w$ regulates the DTW alignment within each window, as introduced in \S~\ref{sec: basics_dtw}. 

The Sakoe-Chiba constraint follows the implementation in the \texttt{dtaidistance} Python package \citep{dtaidistance} and limits the number of bins in one light curve that can be matched to a single bin in another, thereby controlling the degree of temporal compression. 
This Sakoe-Chiba band-type constraint requires that the indices in the two time series satisfy $|i-j| < 2w_w + |n-m|$, thereby restricting the admissible region of $W \in \mathcal{A}$.
For time series of equal length ($n=m$), this reduces to $|i-j| < 2w_w$.
This implicitly limits the number of times a single bin in one light curve can be connected to different bins in the other light curve (also defined as multiplicity).
Indeed, still for the case of $n=m$, for a fixed index $j$, the constraint restricts the admissible indices $i$ to the interval $[j-2w_w+1, j+2w_w-1]$, which contains at most $4w_w-1$ values.
Since each pair $(i,j)$ can appear at most once in the warping path due to the monotonicity condition, and the band constraint restricts the admissible indices, a given bin can be aligned no more than $4w_w-1$ times.
We explored $w_w = 1, 2, 4, 8, 20$, as well as the unconstrained case, in order to assess the impact of alignment flexibility on the resulting similarity distributions. 
Note that $w_w = 1$ is equivalent to the Euclidean distance case, where the alignment between light curves is strictly one-to-one (no warping).
As $w_w$ increases, the alignment becomes more flexible. For intermediate values (e.g., $w_w = 4$, $8$), the results remain stable, improving the separation in $\sigma_{\rm eq}$ between the candidates and control cases (see Fig.~\ref{fig: ECDFs_comparison}) while allowing moderate temporal distortions. 

For larger values, such as $w_w = 20$ or in the unconstrained case, the alignment becomes excessively flexible. Although the resulting $\sigma_{\rm eq}$ values remain consistent, the multiplicities become very large, implying that single bins may be matched to many bins simultaneously. In this regime, the DTW distance is artificially reduced by highly degenerate alignments, compromising the physical interpretability of the comparison.

From a physical perspective, large multiplicities are not meaningful, as they imply that a single temporal feature may correspond to multiple independent features in the other light curve. 
Also, very small multiplicities (towards the Euclidean distance) would imply that large similarities would require strict correlation, which is physically implausible across different systems.

Therefore, the choice of $w_w$ must balance alignment flexibility with a physically plausible mapping between time bins.
Given this trade-off, we adopt $w_w = 4$. This choice preserves the separation between candidate and control cases, as shown in Fig.~\ref{fig: ECDFs_comparison}, while keeping multiplicities moderate, with a maximum value of 7 seen across all analyzed window alignments, thereby maintaining physical interpretability.

\section{Examples of window DTW matching}
\label{appendix: win-matching}

\begin{figure*}
  \centering
    \includegraphics[width=1\columnwidth]{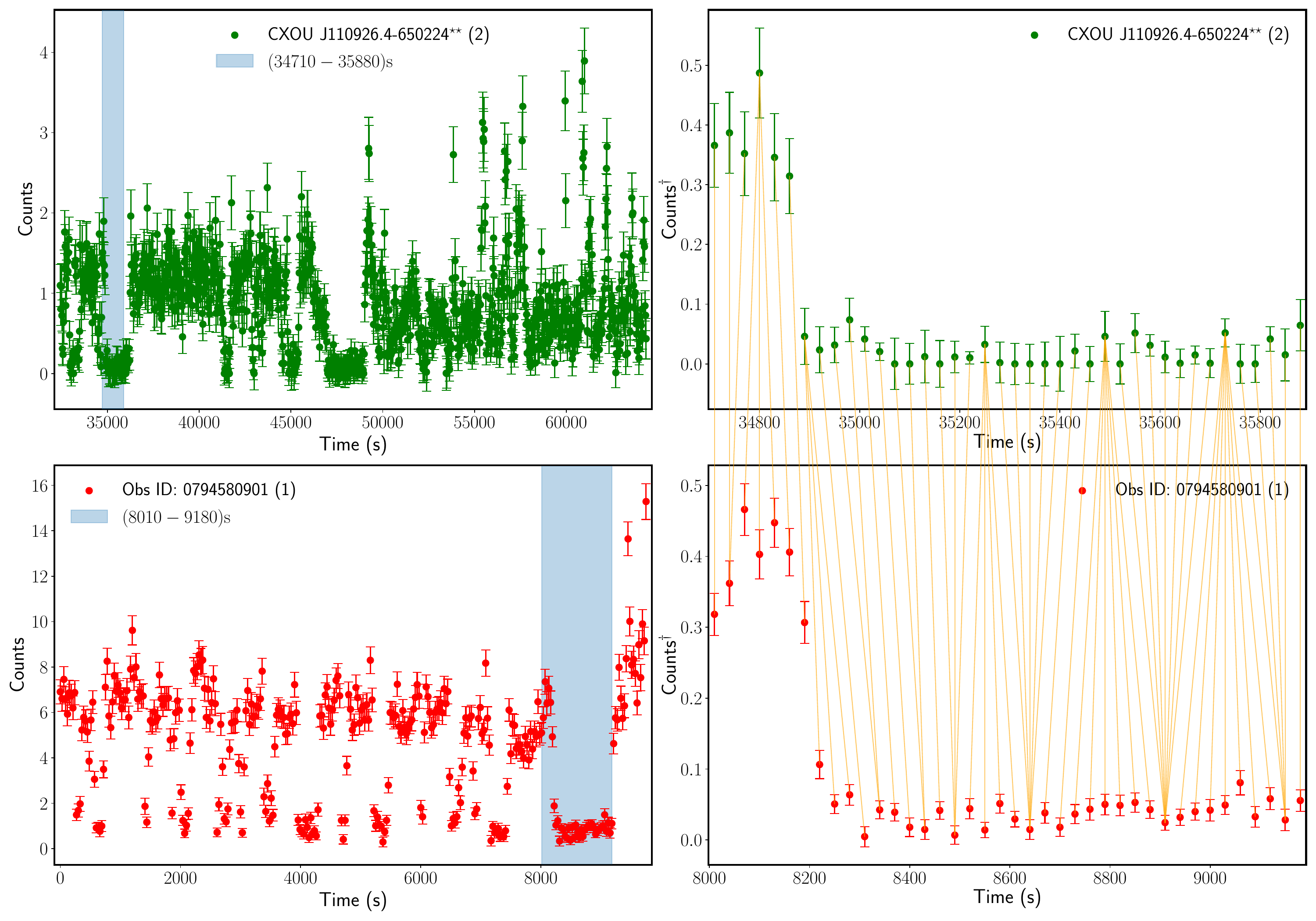}
    \includegraphics[width=1\columnwidth]{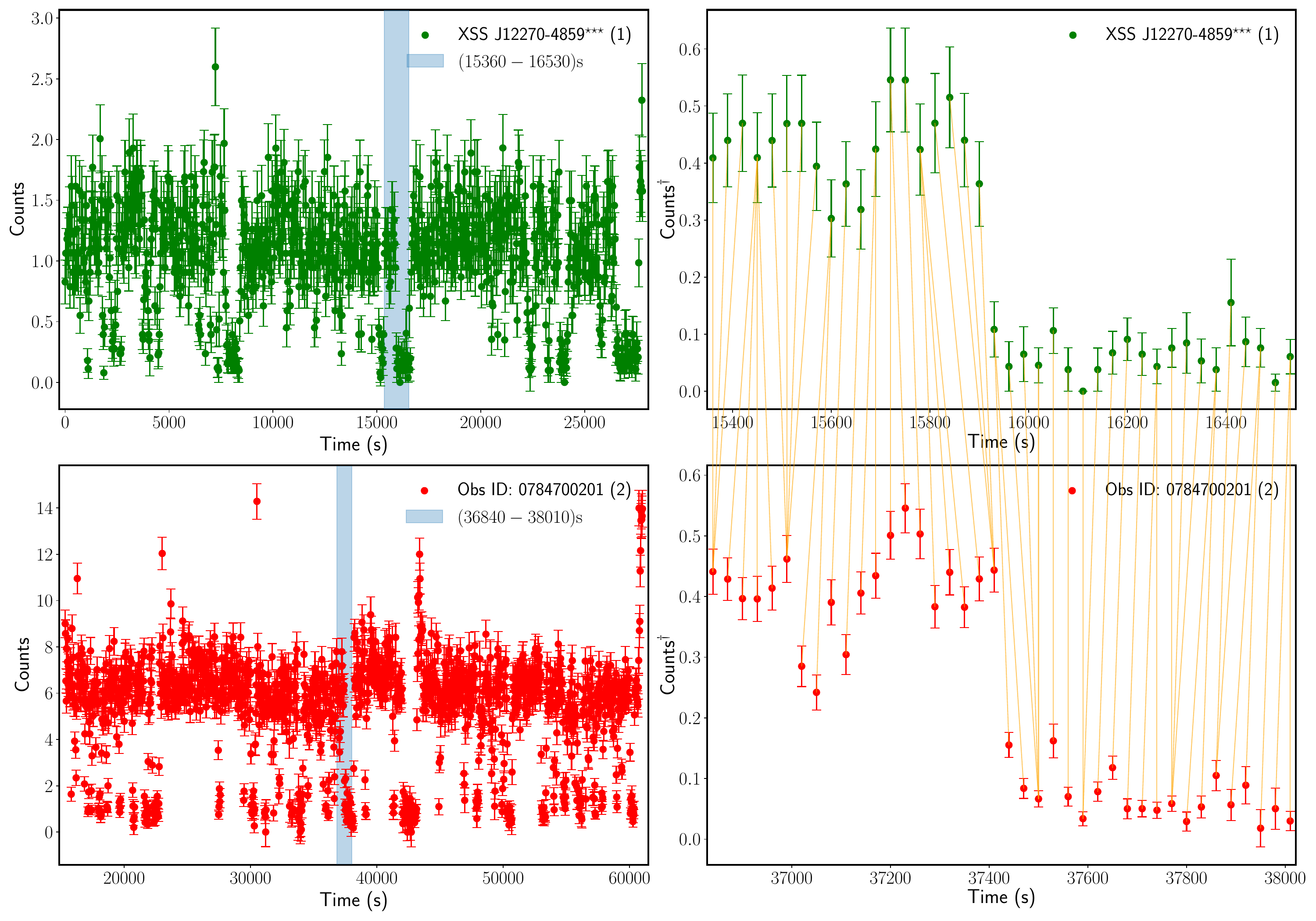}
    \includegraphics[width=1\columnwidth]{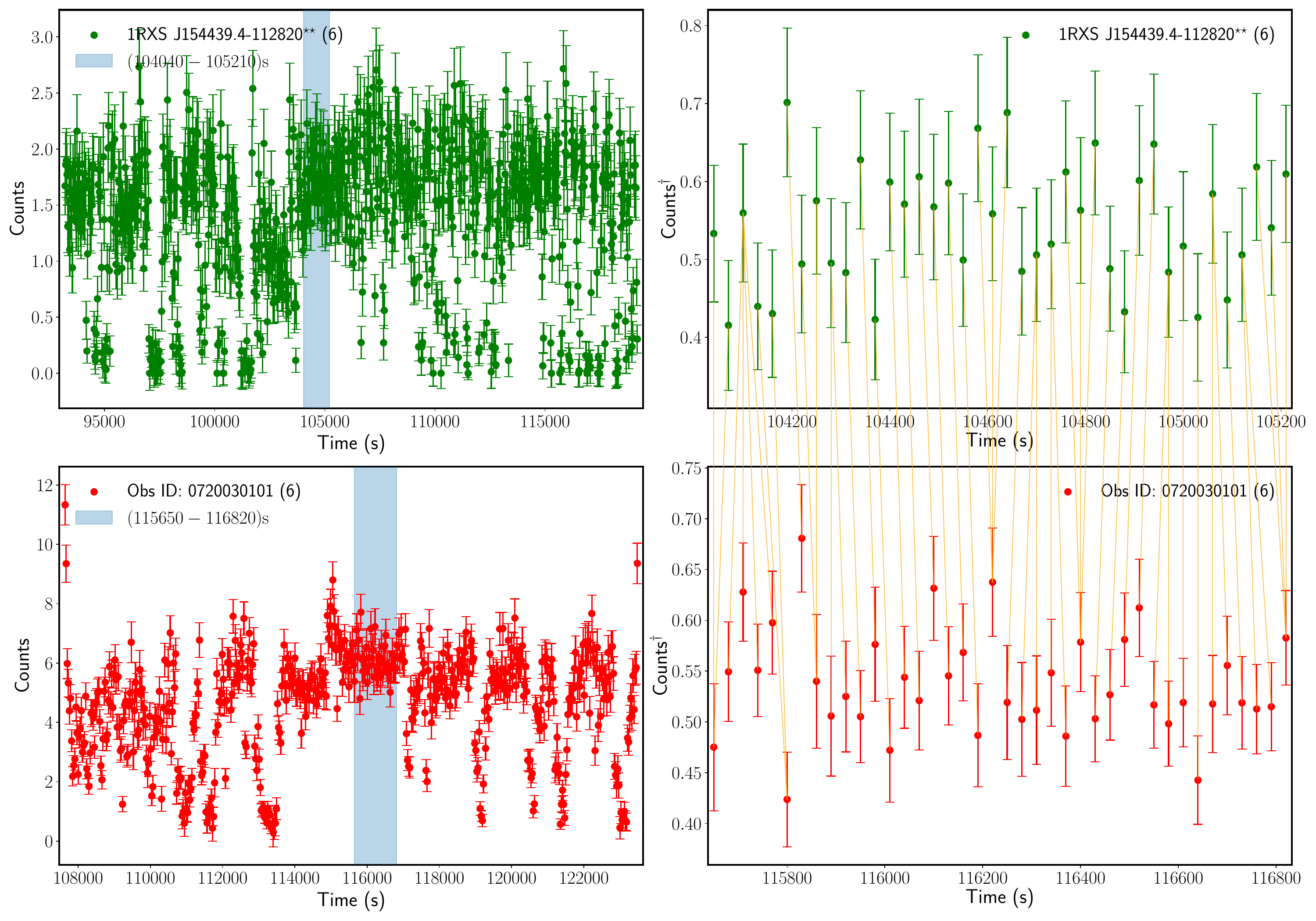}
    \includegraphics[width=1\columnwidth]{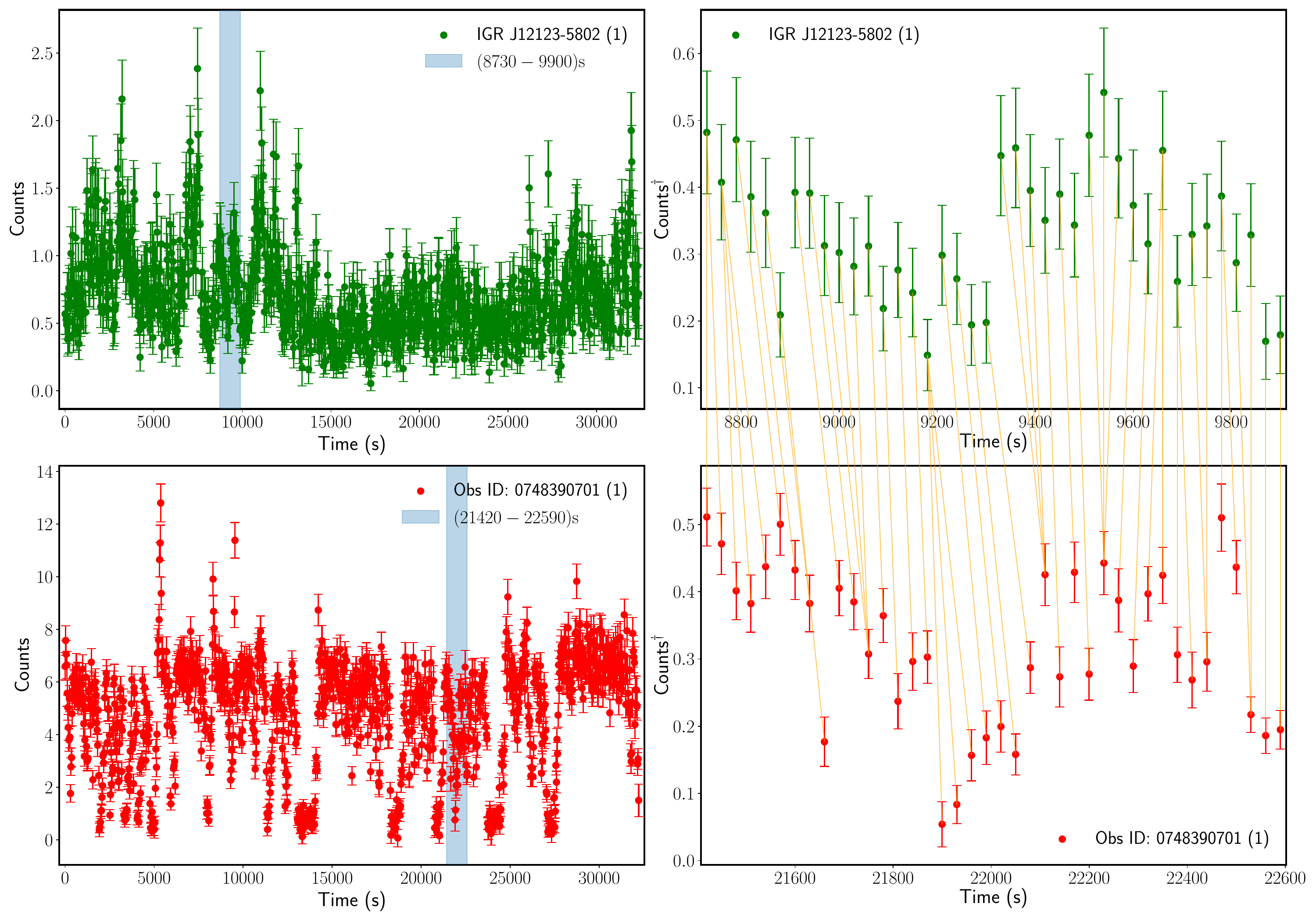}
    \caption{Examples of optimal DTW matches for windows corresponding to light curves of different systems. 
    J1023 appears in red, while the systems it is compared to appear in green. 
    The specific observations are given in the legend. 
    The shaded region indicates the selected bins, accounting for the bin width (30 s) and corresponding to a total temporal coverage of 1200 s (40 bins).
    The orange lines show the DTW alignment between bins, as given by Eq.~(\ref{eq: DTW_optimized}). 
    }
  \label{fig: best_match_windows}
\end{figure*}

We illustrate the DTW alignment behavior at the window level (Eq.~\ref{eq: DTW_optimized}), presenting a set of examples corresponding to the lowest DTW values across different sources.
These examples highlight pairs of windows exhibiting strong morphological similarity despite originating from independent observations of different systems.
See Fig. \ref{fig: best_match_windows}.

\end{appendix}

\end{document}